\documentclass[journal]{IEEEtran}
\ifCLASSINFOpdf
\else
 \usepackage[dvips]{graphicx}
\fi
\usepackage{url}
\usepackage{graphicx}
\usepackage{cite}
\usepackage{amsmath,amssymb,amsfonts}
\usepackage{algorithmic}
\usepackage{graphicx}
\usepackage{textcomp}
\usepackage{xcolor}
\usepackage{bm}
\usepackage{color}
\usepackage{subfigure}
\usepackage{makecell}
\usepackage{algorithm}
\begin{document}
\title{Flexible Intelligent Metasurface for\\Enhancing Multi-Target Wireless Sensing}
\author{Zihao Teng, \IEEEmembership{Graduate Student Member, IEEE}, Jiancheng An, \IEEEmembership{Member, IEEE}, Lu Gan, \IEEEmembership{Member, IEEE}, \\Naofal Al-Dhahir, \IEEEmembership{Fellow, IEEE}, and Zhu Han, \IEEEmembership{Fellow, IEEE}
\thanks{This work is partially supported by NSF ECCS-2302469, CMMI-2222810, Toyota. Amazon, Japan Science and Technology Agency (JST) Adopting Sustainable Partnerships for Innovative Research Ecosystem (ASPIRE) JPMJAP2326, and National Natural Science Foundation of China 62471096.}
\thanks{Z. Teng and L. Gan are with the School of Information and Communication Engineering, University of Electronic Science and Technology of China (UESTC), Chengdu, 611731, China. (e-mail: tzhuestc@163.com; ganlu@uestc.edu.cn).}
\thanks{J. An is with the School of Electrical and Electronics Engineering, Nanyang Technological University (NTU), Singapore 639798 (e-mail: jiancheng.an@ntu.edu.sg).}
\thanks{N. Al-Dhahir is with the Department of Electrical and Computer Engineering, The University of Texas at Dallas, Richardson, TX 75080 USA (e-mail: aldhahir@utdallas.edu).}
\thanks{Z. Han is with the Department of Electrical and Computer Engineering at the University of Houston, Houston, TX 77004 USA, and also with the Department of Computer Science and Engineering, Kyung Hee University, Seoul, South Korea, 446-701 (e-mails: hanzhu22@gmail.com).}}
\markboth{DRAFT}{DRAFT}
\maketitle
\begin{abstract}
Flexible intelligent metasurface (FIM) has emerged as a transformative technology to enhance wireless sensing by dynamically morphing its three-dimensional (3D) surface shape and electromagnetic response. Unlike conventional rigid arrays, an FIM consists of low-cost radiating elements that can independently adjust their positions and radiation characteristics, thereby allowing for real-time optimization of the sensing environment. This paper investigates the impact of FIM on wireless sensing performance. Specifically, we focus on the maximization of the cumulated power of the probing signals at the target locations under the per-antenna power constraint by jointly optimizing the transmit covariance matrix and the surface shape of the transmitting FIM. We propose a block coordinate descend (BCD) algorithm to find a locally optimal solution, by alternatively updating the FIM surface shape and the transmit covariance matrix, while keeping the other one fixed at each step. Furthermore, we analyze the computational complexity and convergence properties of the proposed algorithm and demonstrate that FIM enhances wireless sensing by providing a new design degree-of-freedom to coordinate the correlation between steering vectors at different angles. Numerical results demonstrate that FIM significantly improves wireless sensing performance under the considered multi-target scenario.
\end{abstract}
\IEEEpeerreviewmaketitle
\section{Introduction} 
\IEEEPARstart{I}{n} the sixth generation (6G) wireless networks, the sensing function will become a core driving force for designing innovative technologies. By applying sensing functions, 6G networks will be able to support advanced applications such as high-precision positioning, environmental sensing, and object recognition. These capabilities will enable the development of key technologies like autonomous driving, intelligent manufacturing, and smart cities \cite{you2021towards}. For example, in autonomous driving, high-precision sensing technology can enhance a vehicle's understanding of its surroundings, thus improving safety and efficiency \cite{begum2024auto}. Moreover, the sensing functions of 6G will promote the widespread adoption of virtual reality and augmented reality technologies, providing a more immersive user experience \cite{SP_2025_Teng_Frequency, IoTJ_2025_Teng_Dynamic}. 

Additionally, metasurfaces play a crucial role in the development of 6G wireless networks. The key feature of metasurfaces is their ability to manipulate the behavior of electromagnetic waves through the specific design of unit cell structures, enabling performance gains that natural materials cannot achieve \cite{di2020smart}. One of the core applications of metasurfaces in 6G networks is the deployment of reconfigurable intelligent surfaces (RISs). By precisely managing the amplitude and phase of incident electromagnetic waves, RIS can enhance signal coverage, increase data rates, and reduce interference without additional energy consumption \cite{an2024codebook,di2020smart}. Building on this capability, the authors of \cite{buz2022fon} considered the fundamental problem of target detection in a RIS-aided multiple-input multiple-output (MIMO) radar. Specifically, they showed the benefits of the RISs and shed light on the interplay among the key system parameters. The authors of \cite{song2023crb} investigated RIS-enabled non-line-of-sight (NLoS) wireless sensing based on Cramér-Rao bound optimization. Furthermore, Lyu \emph{et al.} investigated RIS-aided millimeter-wave (mmWave) integrated sensing and communications (ISAC) systems \cite{lv2024noma}. Additionally, \cite{wangwcnc2024} applied RISs and passive radars to ISAC systems for enhancing localization and detection performance, and \cite{baz2025jsac} applied RISs to bistatic ISAC systems for mitigating multipath interference.

Expanding upon RIS, stacked intelligent metasurfaces (SIM) have been proposed, which are comprised of multiple programmable transmissive metasurfaces \cite{TAP_2025_An_Emerging, an2023stacked}. Hence, SIM can perform advanced signal processing tasks in the wave domain, such as ISAC \cite{niu2024isac}, MIMO precoding \cite{an2023stacked2}, and direction-of-arrival (DoA) estimation \cite{an2024two}. However, all the aforementioned metasurfaces have generally adopted rigid metamaterials, which impose limitations on the degrees of freedom (DoF) in adjusting their configuration.

Thanks to recent advances in micro/nanofabrication, it has become possible to create flexible intelligent metasurfaces (FIMs) on pliable substrates such as polydimethylsiloxane (PDMS) \cite{kamali2016decoupling}. These materials allow metasurfaces to not only control electromagnetic (EM) waves but also physically morph their substrate\footnote{For a demonstration of the real-time morphing capability of an FIM, please visit {https://www.eurekalert.org/multimedia/950133}.}, enabling a new type of devices that can adapt to various conditions, such as changes in curvature or environmental demands \cite{bai2022dynamically}. This breakthrough opens up the potential for FIMs in various fields, including soft robotics, wearable sensors, and implantable devices \cite{kamali2016decoupling}. While recent studies \cite{anfim2025, arXiv_2025_An_Flexible} have investigated the applications of FIM in enhancing the channel gain in wireless communication systems, the use of FIM for wireless sensing remains an unexplored research area. In particular, FIMs are well-suited for wireless sensing. Their morphing ability provides additional DoF, enabling FIMs to coordinate the correlation between steering vectors in different directions. In this paper, we investigate enhanced MIMO sensing with an FIM by jointly optimizing the transmit covariance matrix and the three-dimensional (3D) surface shape of the transmitting FIM. Note that recent fluid antenna systems (FAS) and movable antenna systems (MAS) also aim to improve sensing performance by physically reconfiguring the antenna geometry \cite{zhou2024fluid,ma2024ma}. However, these systems typically rely on discrete mechanical repositioning of antenna elements, which is limited by the response speed of motor-driven components. In contrast, firstly, FIM enables fast and continuous surface reconfiguration through Lorentz-force-driven deformation \cite{bai2022dynamically}. Secondly, FIM supports dense array integration and flexible deployment on curved or lightweight platforms, unlike the bulky structures required by FAS/MAS.

More specifically, the main contributions of this paper are summarized as follows:
\begin{enumerate}
	\item We formulate the FIM-enhanced MIMO wireless sensing problem to maximize probing signal power at target locations by jointly optimizing the transmit covariance matrix and the 3D surface shape of the transmitting FIM.
	\item We propose an efficient block coordinate descend (BCD) algorithm to address the non-convexity of the joint optimization problem. This algorithm iteratively optimizes one of the transmit covariance matrix and the surface shape of the transmitting FIM with the other one being held fixed, yielding a locally optimal solution.
	\item Simulation results show that our proposed sensing waveform design with FIM can significantly outperform that with the traditional rigid arrays and other benchmark schemes in terms of the probing signal power at target locations.
\end{enumerate}

\section{Wireless Sensing Model}
\begin{figure}[]
	\centering
	\includegraphics[width=0.5\textwidth]{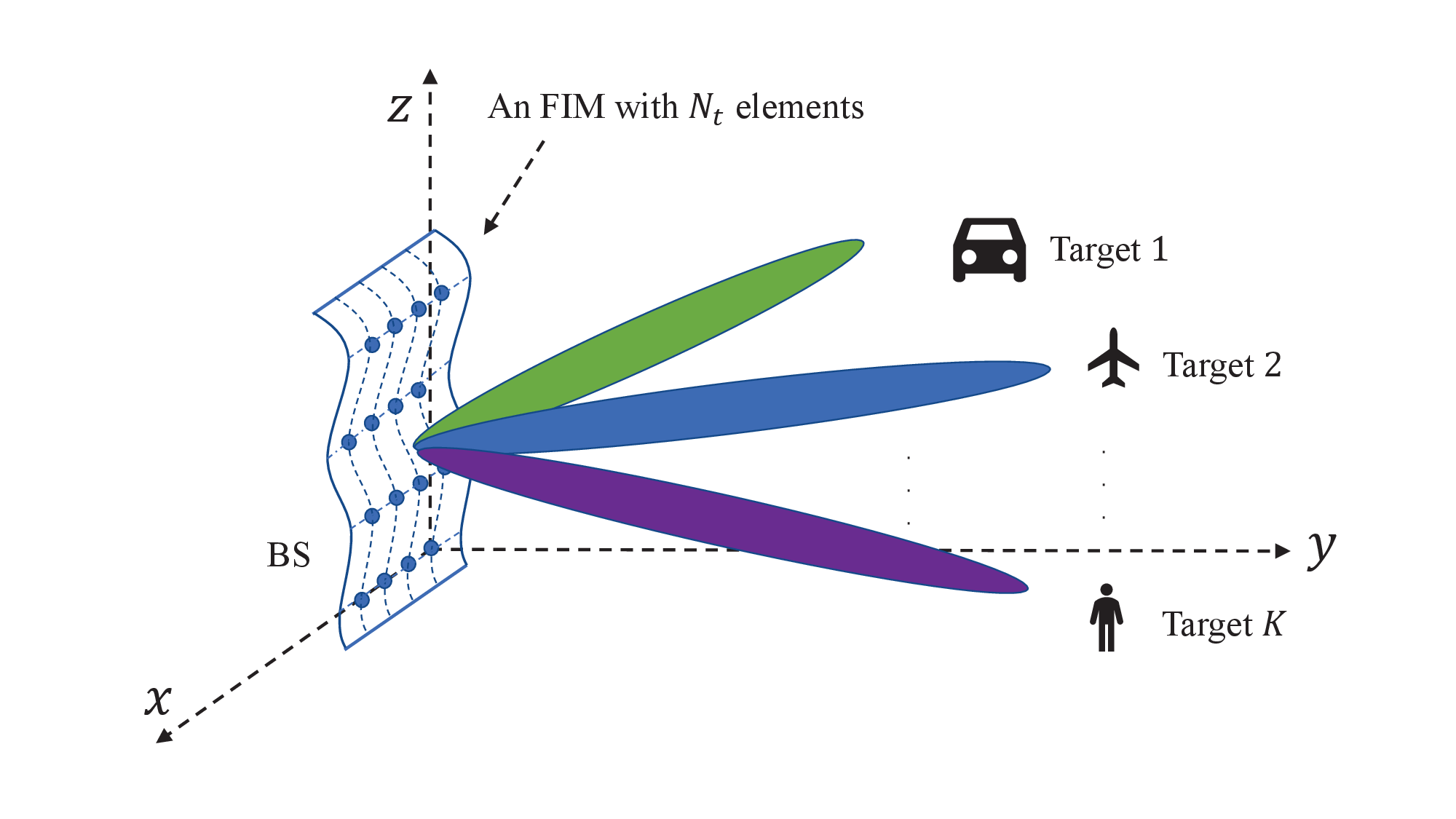}
	\caption{The illustration of an FIM-aided MIMO wireless sensing system.}
	\label{illustration}
\end{figure}
We consider a MIMO sensing system, as illustrated in Fig. \ref{illustration}, where the base station (BS) is equipped with an FIM operating in the full-duplex mode. In particular, the self-interference can be mitigated through passive or active techniques \cite{zhang2015duplex}. The antenna arrays on the FIM are modeled as a flexible uniform planar array (UPA) on the $xz$ plane. In contrast to conventional rigid arrays, the FIM is capable of morphing its surface shape \cite{bai2022dynamically}. The deformation of each element is driven by a magnetic field that is applied parallel to the plane of the surface. As a result, the induced Lorentz force acts perpendicular to the surface, allowing the position of each element to be independently adjusted along the normal direction (i.e., $y$-axis direction) of the FIM. The total number of transmitting antenna elements is $N_t = N_{t1} \times N_{t2}$, where $N_{t1}$ and $N_{t2}$ denote the number of elements along the $x$-axis and $z$-axis, respectively. 

We assume that there are $K $ targets located at $\left\{( {{\theta _k},{\varphi _k}} )\right\}_{k = 1}^K$, where $\theta_k\in [0, \pi]$ and $\varphi_k\in [0, \pi]$ represent the elevation and azimuth angles of the $k$th target, respectively. The transmitting steering vector ${\bf{a}}({\theta _k},{\varphi _k},\Delta {{\bf{d}}^t})\in \mathbb{C}^{{N_t} \times 1}$ can be expressed as
\begin{align}
	{\bf{a}}({\theta _k},{\varphi _k},\Delta {{\bf{d}}^t}) 
	= \left [ {{\bf{a}}_z}({\theta _k}) \otimes {{\bf{a}}_x}({\theta _k},{\varphi _k}) \right ] \odot {{\bf{a}}_y}({\theta _k},{\varphi _k},\Delta {{\bf{d}}^t}),
\end{align}
for $k=1,2,\cdots,K$, where ${{\bf{a}}_x}({\theta _k},{\varphi _k}) \in \mathbb{C}^{{N_{t1}} \times 1}$, ${{\bf{a}}_z}({\theta _k}) \in \mathbb{C}^{{N_{t2}} \times 1}$ are defined by
\begin{align}
		{{\bf{a}}_x}({\theta _k},{\varphi _k}) &\triangleq \left[1,e^{ - j{k_c}v_{x,k}}, \cdots,e^{ - j{k_c}({N_{t1}} - 1)v_{x,k}} \right]^T,\\
		{{\bf{a}}_z}(\theta_k ) &\triangleq \left[ 1,{e^{ - j{k_c}v_{z,k} }},	\cdots, {e^{ - j{k_c}({N_{t2}} - 1)v_{z,k} }} \right]^T,
\end{align}
with $v_{x,k}=d_x^t\sin {\theta _k}\cos {\varphi _k}$, $v_{z,k}=d_z^t\cos \theta_k$. Moreover, $d_x^t$ and $d_z^t$ denote the spacing between adjacent antenna elements in the $x$- and $z$-directions of the transmitting FIM, respectively. The parameter $k_c=2\pi/\lambda$ represents the wavenumber, with $\lambda$ denoting the carrier’s wavelength. Furthermore, let $\Delta {{\bf{d}}^t} = {\left[\Delta d_1^t,\Delta d_2^t,...,\Delta d_{{N_t}}^t\right]^T} \in \mathbb{R}^{N_t \times 1}$ denote the $y$-coordinate of each transmitting FIM element, we have
\begin{align}
	\begin{split}
		{{\bf{a}}_y}(\theta_k ,\varphi_k ,\Delta {{\bf{d}}^t}) &= \left[e^{ - j{k_c}\Delta d_1^t\sin \theta_k \sin \varphi_k}, e^{ - j{k_c}\Delta d_2^t\sin \theta_k \sin \varphi_k }, \right.\\
		&\left.	\cdots,e^{ - j{k_c}\Delta d_{{N_t}}^t\sin \theta_k \sin \varphi_k }\right]^T \in \mathbb{C}^{{N_{t}} \times 1}.
	\end{split}
\end{align}
Due to the elastic limitation of the surface material, we have a maximum morphing range $d_{\max }^t$, and the deformation distance should satisfy $\left| {\Delta d_{{n_t}}^t} \right| \le d_{\max }^t,{n_t} = 1,2,...,{N_t}$. 

As a result, let ${\bf{X}} = \left[ {{\bf{x}}(1),{\bf{x}}(2),...,{\bf{x}}(T)} \right] \in \mathbb{C}^{N_t \times T}$ denote the transmitting waveform matrix, with $T$ being the number of snapshots. We assume that the target location remains approximately constant during the whole snapshots \cite{stoica2007prob, tsp-2008-li-crb, song2023crb}. The received data matrix ${\bf{Y}} \in \mathbb{C}^{{N_t} \times T}$ can be written as
\begin{align}
	{\bf{Y}} &= \sum\limits_{k = 1}^K {{\alpha _k}} {\bf{a}}^*({\theta _k},{\varphi _k},\Delta {{\bf{d}}^t}){{\bf{a}}^H}({\theta _k},{\varphi _k},\Delta {{\bf{d}}^t}){\bf{X}} + {\bf{N}}\nonumber\\
	&= {\bf{A}}^*{\bf{\Lambda }}{{\bf{A}}^H}{\bf{X}} + {\bf{N}},
\end{align}
where ${\bf{A}} \in \mathbb{C}^{{N_t} \times K}$ denotes the array's response matrix 
\begin{align}
	{\bf{A}} = \left[{\bf{a}}({\theta _1},{\varphi _1},\Delta {{\bf{d}}^t}),{\bf{a}}({\theta _2},{\varphi _2},\Delta {{\bf{d}}^t}), \cdots,{\bf{a}}({\theta _K},{\varphi _K},\Delta {{\bf{d}}^t})\right],
\end{align}
${\bf{\Lambda }} = {\text{ diag}}({\bm{\alpha }})\in \mathbb{C}^{{K} \times K}$, $\bm{\alpha }=\left[{\alpha _1},{\alpha _2},\cdots,{\alpha _K}\right]^T \in \mathbb{C}^{{K} \times 1}$, $\left\{ {{\alpha _k}} \right\}_{k = 1}^K$ are the complex-valued channel coefficients dependent on the targets’ radar cross section (RCS). The matrix ${\bf{N}} \in \mathbb{C}^{N_t \times T}$ denotes the additive white Gaussian noise, satisfying ${\text{vec}}({\bf{N}})\sim{\mathcal{CN}}(0,\sigma _n^2{{\bf{I}}_{{N_t}T}})$, with $\sigma_n^2$ representing the noise power. Therefore, the cumulated power of the probing signals at the target locations is given by
\begin{align}
	P_c=\sum\limits_{k = 1}^K {{{\bf{a}}^H}({\theta _k},{\varphi _k},\Delta {{\bf{d}}^t}){{\bf{R}}_{\text{X}}}{\bf{a}}({\theta _k},{\varphi _k},\Delta {{\bf{d}}^t})}=\text{tr}\left({\bf{R}}_\text{X}{\bf{B}}\right),
\end{align}
where ${{\bf{R}}_\text{X}} = \frac{1}{T}{\bf{X}}{{\bf{X}}^H}\in\mathbb{C}^{{N_{t}} \times {N_{t}}}$
denotes the transmit covariance matrix, ${\bf{B}}\in\mathbb{C}^{{N_{t}} \times {N_{t}}}$ is
\begin{align}\label{BB}
	{\bf{B}} = \sum\limits_{k = 1}^K {{\bf{a}}({\theta _k},{\varphi _k},\Delta {{\bf{d}}^t}){{\bf{a}}^H}({\theta _k},{\varphi _k},\Delta {{\bf{d}}^t})}={\bf{A}}{\bf{A}}^H. 
\end{align}

\section{Power Maximization for FIM-aided Wireless Sensing}

\subsection{Problem Formulation}
In this paper, we aim to maximize the probing signals' power at the target locations by jointly optimizing the surface shape $\Delta {{\bf{d}}^t}$ of the transmitting FIM within the maximum morphing range $d_{\max }^t$, and the transmit covariance matrix ${\bf{R}}_\text{X}$ under the per-antenna power constraint $P_t/N_t$. Specifically, the joint optimization problem is formulated as
\begin{subequations}\label{prob1}
	\begin{align}
		\mathop {\max }\limits_{{{\bf{R}}_\text{X}},\Delta {{\bf{d}}^t}}\quad& P_c\triangleq\text{tr}\left({\bf{R}}_\text{X}{\bf{B}}\right)\\
		\text{s.t.}\quad\quad& \text{diag}\left({\bf{R}}_\text{X}\right) = \frac{P_t}{N_t},\label{rx1}\\
		& {{\bf{R}}_\text{X}} \succeq {\bf{0}},\label{rx2}\\
		&\left| {\Delta d_{{n_t}}^t} \right| \le d_{\max }^t,\quad{n_t} = 1,2,...,{N_t}\label{rd1},
	\end{align}
\end{subequations}
where \eqref{rx2} means that ${{\bf{R}}_\text{X}}$ is a positive semi-definite matrix. It is challenging to solve problem \eqref{prob1} directly since the objective function is non-convex related to the FIM's surface shape. To efficiently tackle this non-convex problem, in the next subsection, we propose a BCD algorithm that decomposes \eqref{prob1} into two tractable subproblems, namely the optimization of the transmit covariance matrix ${\bf{R}}_\text{X}$ and the surface-shape morphing of the transmitting FIM. For each subproblem, we focus on one set of variables while keeping the other one fixed.

\subsection{The Proposed BCD Algorithm}
\subsubsection{Optimize ${\bf{R}}_\text{X}$ with a Given FIM's Shape ${\Delta {{\bf{d}}^t}}$}
Here, we aim to optimize the waveform covariance matrix ${\bf{R}}_\text{X}$ with a given transmitting FIM's shape ${\Delta {{\bf{d}}^t}}$. Therefore, the subproblem can be expressed as
\begin{subequations}\label{rxtrans}
	\begin{align}
		\mathop {\max }\limits_{{{\bf{R}}_\text{X}}}\quad& \text{tr}\left({\bf{R}}_\text{X}{\bf{B}}\right)\\
		\text{s.t.}\quad \ & \eqref{rx1}, \eqref{rx2}.
	\end{align}
\end{subequations}
Note that problem \eqref{rxtrans} is reduced to a semi-definite program (SDP), which is convex and can be optimally addressed using convex optimization tools, e.g., the SDPT3 solver in the CVX toolbox \cite{cvx}.
\subsubsection{Optimize FIM's Shape $\Delta {{\bf{d}}^t}$ with a Given ${\bf{R}}_\text{X}$}
Next, we aim to optimize the transmitting FIM's surface shape $\Delta {{\bf{d}}^t}$ with a given waveform covariance matrix ${\bf{R}}_\text{X}$. In this case, the FIM's shape optimization problem is formulated as
\begin{subequations}\label{rxgivenR}
	\begin{align}
		\mathop {\max }\limits_{\Delta {{\bf{d}}^t}}\quad&P_c\triangleq \text{tr}\left({\bf{R}}_\text{X}{\bf{B}}\right)\\
		\text{ s.t.}\quad \ &\eqref{rd1}.
	\end{align}
\end{subequations}
The objective function in \eqref{rxgivenR} is highly non-convex with respect to (\emph{w.r.t.}) $\Delta {{\bf{d}}^t}$, making it challenging to obtain the globally optimal solution. To address this issue, we employ the gradient descent algorithm to find a suboptimal solution. Starting with an initial surface shape of the transmitting FIM, we iteratively adjust the surface shape by following the direction of the gradient to gradually increase the probing signal power at the target locations. Specifically, the surface shape of the transmitting FIM is updated by
\begin{align}\label{gra_decentdt}
	\Delta {{\bf{d}}^t}\leftarrow\Delta {{\bf{d}}^t}+\varepsilon\nabla_{\Delta {{\bf{d}}^t}}P_c,
\end{align}
where $ \varepsilon > 0$ represents the step size. Notice that due to the fact that the exact Lipschitz constant of the gradient is difficult to compute in a closed form, the selection of an appropriate step size is done using a backtracking line search technique to prevent overshooting. Specifically, an acceptable $ \varepsilon $ should satisfy Armijo conditions. Moreover, $\nabla_{\Delta {{\bf{d}}^t}}P_c$ denotes the gradients of the objective function \emph{w.r.t.} the surface-shape configurations of $\Delta {{\bf{d}}^t}$. The detailed calculation of $\nabla_{\Delta {{\bf{d}}^t}}P_c$ can be found in Appendix A.

Moreover, to satisfy the maximum morphing range constraints in \eqref{rd1}, a projection function is utilized to restrain the FIM elements given by \eqref{gra_decentdt} within their allowable ranges
\begin{align}
	\Delta d_{{n_t}}^t &= \text{sgn}\left( {\Delta d_{{n_t}}^t} \right)\min \left( {\left| {\Delta d_{{n_t}}^t} \right|, d_{\max }^t} \right), n_t = 1,2,\cdots,N_t,\label{rangedt}
\end{align}
where $\text{sgn} (x) $ denotes the sign function.

As a result, the proposed BCD algorithm is summarized as Algorithm 1.
\begin{algorithm}[]	
	\caption{
		The Proposed BCD Algorithm for Power Maximization
	}
	\label{Algorithm3}
	{\begin{tabular}{l l}	
			\textbf{Input:} $ \theta_k $, $ \varphi_k $, $k=1,2,\cdots, K$, $P_t$, $\lambda$, $N_{t1}$, $N_{t2}$, $d_x^t$, $d_z^t$,\\ $d_{\max}^t$, $\zeta$, $S$.\\
			1: Initialize the surface shape $\Delta {{\bf{d}}^t}$ of the transmitting FIM.\\
			2: \textbf{Repeat} \\
			3: \quad Optimize the transmit covariance matrix for the\\
			\quad \ \quad given $\Delta {{\bf{d}}^t}$ by solving problem \eqref{rxtrans}.\\
			4: \ \quad Initialize $\Delta {{\bf{d}}_{(1)}^t}=\Delta {{\bf{d}}^t}$, $s=1$.\\
			5: \ \quad \textbf{While} $\left\|\nabla_{\Delta {{\bf{d}}^t}}P_c\right\|>\zeta$\\
			6: \ \quad \quad Update the transmitting FIM's shape by \eqref{gra_decentdt}.\\
			7: \ \quad \quad Scale the transmitting FIM's shape into\\ 
			\quad \quad \, \quad the allowable morphing range by \eqref{rangedt}.\\
			8: \ \quad \quad $s=s+1$.\\
			9: \ \quad \quad \textbf{If} $s>S$\\
			10: \ \quad \quad \ \ \textbf{break}\\
			11: \quad \quad \textbf{end}\\
			12: \ \quad\textbf{end}\\
			13: \ \quad Update $\Delta {{\bf{d}}^t}=\Delta{{\bf{d}}_{(s)}^t}$.\\
			14: \textbf{Until} the fractional increase of $P_c$ falls below a preset \\
			 \ \ \quad threshold or the maximum tolerable number of iterations \\
			 \ \ \quad is reached.\\
			\textbf{Output:} ${\bf{R}}_\text{X}$, $\Delta {{\bf{d}}^t}$. 
	\end{tabular}}
\end{algorithm}
\subsection{Complexity and Convergence Analysis}

Next, we analyze the computational complexity of the proposed BCD algorithm. In each iteration, the worst-case computational complexity of solving the SDP problem \eqref{rxtrans} is $\mathcal{O}\left( (N_t)^{4.5}\log(1/\varUpsilon)\right)$, given a solution accuracy $\varUpsilon>0$ \cite{2010_sdp_luo}. Furthermore, the complexity of optimizing the surface shape heavily depends on calculating the gradient. From \eqref{BBB}, \eqref{PA}, we note that ${\partial }{\bf{A}}/{{\partial \Delta d_{{n_t}}^t}}$ and ${\partial }{\bf{B}}/{{\partial \Delta d_{{n_t}}^t}}$ are sparse matrices. For a given $n_t$, ${\partial }{\bf{A}}/{{\partial \Delta d_{{n_t}}^t}}$ has non-zero elements only in the $n_t$-th row, and ${\partial }{\bf{B}}/{{\partial \Delta d_{{n_t}}^t}}$ has non-zero elements only in the $n_t$-th row and the $n_t$-th column. Thus, the computational complexity of calculating \eqref{BBB} and \eqref{B1} are $\mathcal{O}\left(KN_t\right)$ and $\mathcal{O}\left(N_t^2\right)$, respectively. By calculating all $N_t$ partial derivatives, the complexity for calculating the gradient $\nabla_{\Delta {{\bf{d}}^t}}P_c$ is $\mathcal{O}\left(N_t^3+KN_t^2\right)$. Since we have $K \ll N_t$, by omitting low-order terms, the overall computational complexity of the proposed algorithm is $\mathcal{O}\left(I\left((N_t)^{4.5}\log(1/\varUpsilon)+I_1N_t^3\right)\right)$, where $I_1$ and $I$ represent the number of iterations for optimizing the surface shape of the transmitting FIM and BCD optimization, respectively.

Furthermore, by selecting an appropriate step size according to the Armijo conditions, the objective function monotonically increases in each iteration step. Moreover, the objective function is upper bounded due to the power constraint in \eqref{rx1}. Thus, the proposed algorithm is guaranteed to converge to at least a local maximum of $P_c$. By leveraging multiple random initializations, we can further improve the performance of Algorithm 1.

\subsection{Further Discussions}
In Section III-A, we consider the signal power maximization problem under the per-antenna power constraint. To gain more insights about the FIM-enhanced wireless sensing, we consider the total power constraint by modifying \eqref{rx1} with $\text{tr}\left({\bf{R}}_\text{X}\right) ={P_t}$.
Under this modification, the waveform covariance matrix ${\bf{R}}_\text{X}$ in \eqref{prob1} has a closed-form solution given by \cite{stoica2007prob}
\begin{align}
{\bf{R}}_\text{X} ={P_t}{\bf{u}}{\bf{u}}^H,
\end{align}
where ${\bf{u}}$ is the unit-norm eigenvector associated with the maximum eigenvalue of ${\bf{B}}$, and the maximum signal power is
\begin{align}\label{pc1}
	P_c ={P_t}\left\| {\bf{A}} \right\|_2^2.
\end{align}
$\left\| {\bf{A}} \right\|_2=\lambda_{\max}({\bf{AA}}^H)=\lambda_{\max}({\bf{B}})$ denotes the spectral norm of ${\bf{A}}$, i.e., the largest eigenvalue of ${\bf{B}}$. Furthermore, let $\lambda _k, k=1,2,\cdots K$ denote the non-zero eigenvalues of ${\bf{B}}$, we have $\text{tr}\left({\bf{B}}\right) ={KN_t}=\sum\limits_{k = 1}^{{K}} {{\lambda _k}} ({\bf{B}})$. Since ${\bf{B}}$ is a positive semi-definite matrix, we have $\lambda _k({\bf{B}})\ge 0$, $k=1,2,\cdots K$. It is easy to conclude that when $\text{rank}({\bf{B}})=1$, \eqref{pc1} can be maximized. Moreover, noting that $\text{rank}({\bf{B}})=\text{rank}({\bf{A}}{\bf{A}}^H)=\text{rank}({\bf{A}})$, which means that each column of ${\bf{A}}$ is linearly correlated. In this case, the signal components associated with multiple targets can be simultaneously maximized.

In practice, the array response matrix ${\bf{A}}$ associated with multiple targets may not satisfy the rank-1 condition. However, thanks to the surface shape morphing capability of the FIM to dynamically adjust the positions of array elements, the wireless sensing performance can be significantly enhanced. This design DoF allows FIM to actively modify the array response matrix ${\bf{A}}$, and increase the correlation between steering vectors across different target directions. As a result, the coherent gain is obtained to improve detection performance by amplifying the cumulated probing signal power in multiple directions.

\section{Simulation Results}
In this section, computer simulations are carried out to evaluate the performance of utilizing FIM to enhance multi-target wireless sensing. As shown in Fig. \ref{illustration}, we consider a MIMO sensing system equipped with a transmitting FIM. The number of transmitting elements along the $x$-axis and $z$-axis direction are $N_{t1}=N_{t2}=10$. Moreover, the system operates at 28 GHz, and the spacing between adjacent antenna elements in the $x$- and $z$-directions of the transmitting FIM is $d_x^t=d_z^t=0.5\lambda$. The target number is $K=3$, which are located at $\theta_1=30^\circ,\varphi_1=60^\circ$, $\theta_2=30^\circ,\varphi_2=120^\circ$, and $\theta_3=135^\circ,\varphi_3=90^\circ$, respectively, with unit RCS. For the gradient descent algorithm, the maximum allowed number of iterations is set to $S=1000$, and the gradient threshold for achieving a locally optimal shape of the transmitting FIM is set to $\zeta=10^{-6}$. For the proposed BCD algorithm, the maximum allowed number of iterations is set to 50, and the convergence threshold in terms of the fractional increase in the objective function is set to $-30$ dB.
\begin{figure*}[]
	\centering
	\subfigure[]{
		\label{raapabp}
		\includegraphics[width=0.4\textwidth]{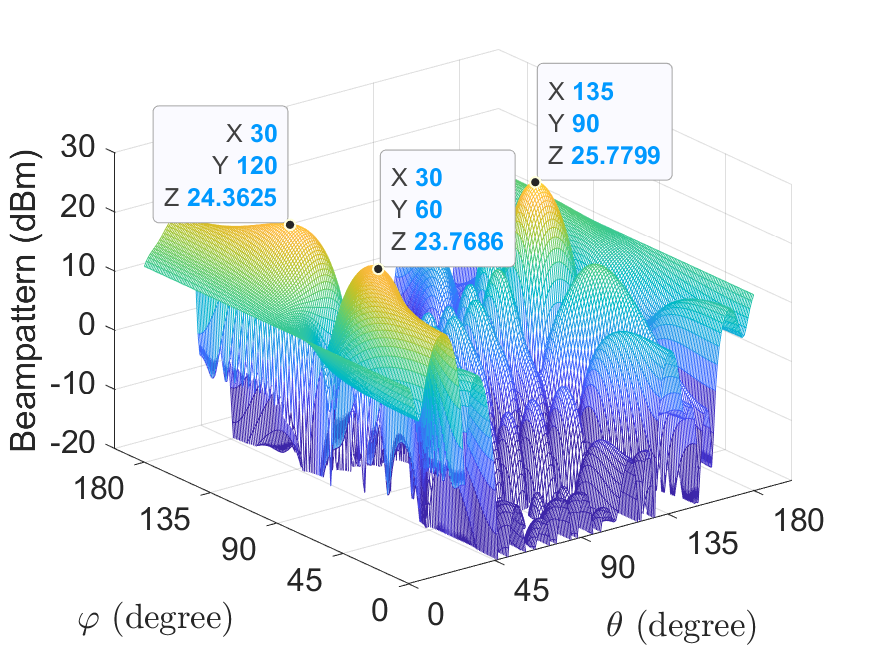}
	}
	\subfigure[]{
		\label{fimpabp}
		\includegraphics[width=0.4\textwidth]{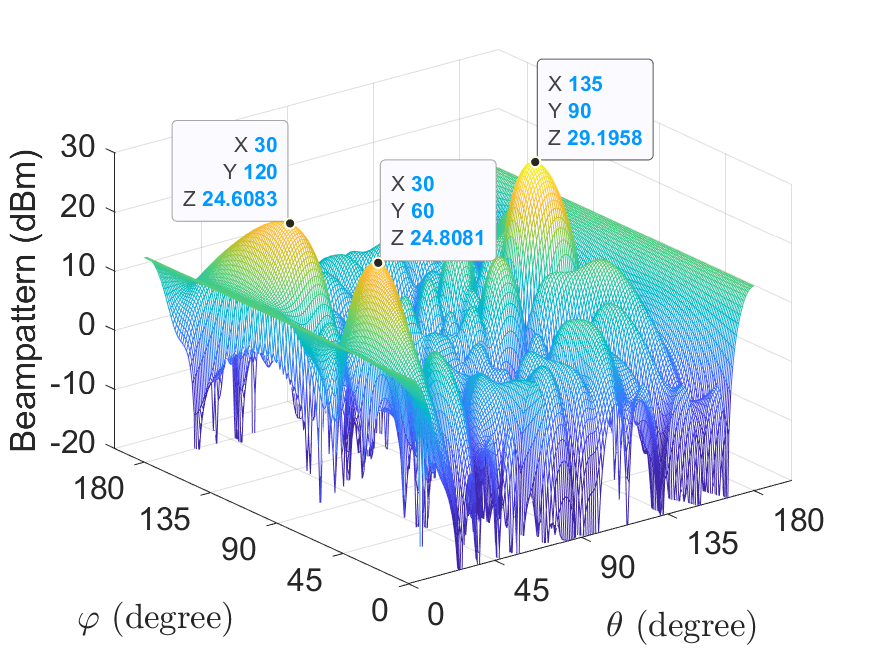}
	}
	\subfigure[]{
		\label{raamimobp}
		\includegraphics[width=0.4\textwidth]{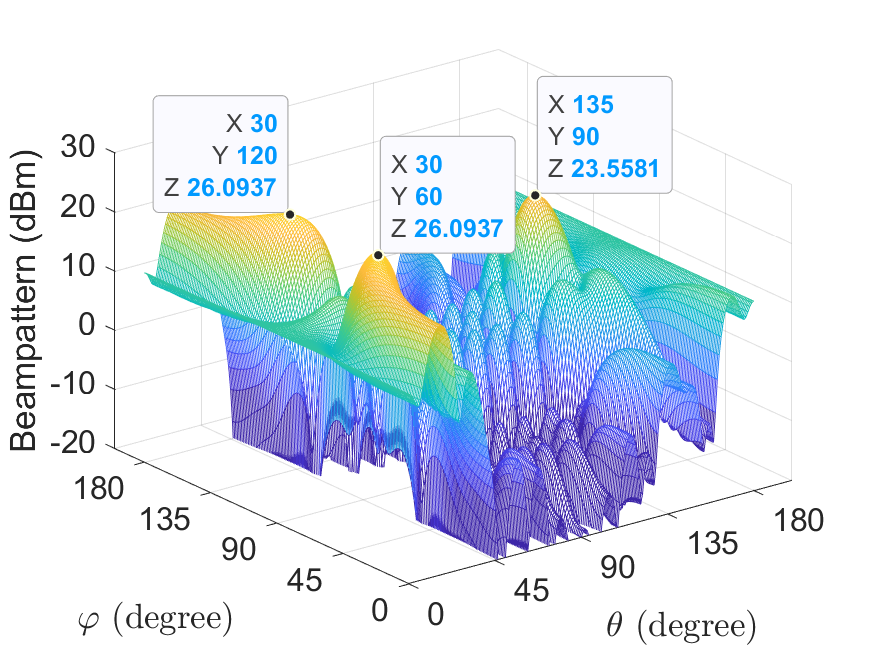}
	}
	\subfigure[]{
		\label{fimmimobp}
		\includegraphics[width=0.4\textwidth]{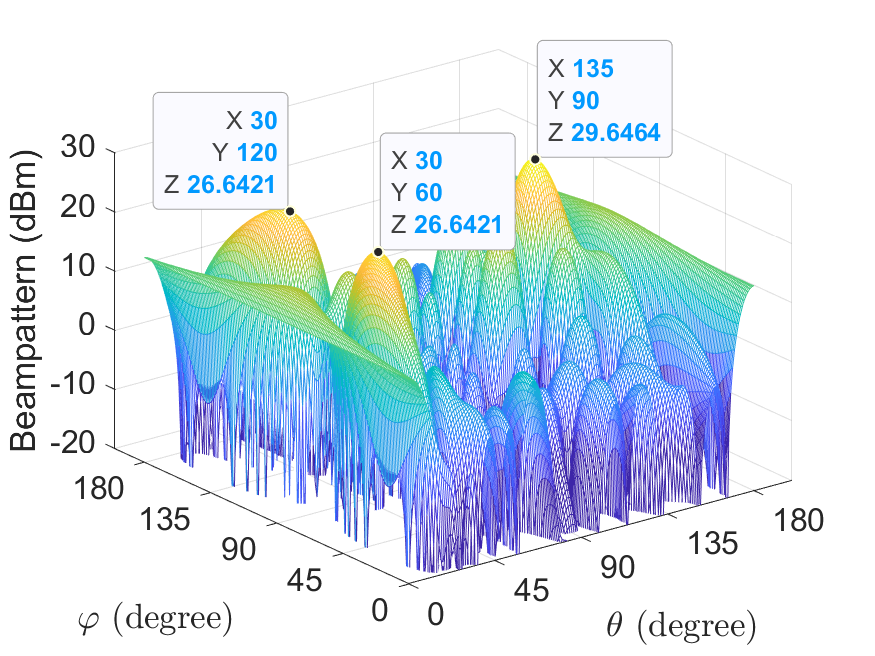}
	}
	\caption{Beampatterns under four different transmitters,
		where: (a) RAA-PA;
		(b) FIM-PA;
		(c) RAA-MIMO;
		(d) FIM-MIMO.}
	\label{bp}
\end{figure*}

\subsection{Beampattern Comparison}
In this subsection, we examine the beampettern to evaluate the enhanced sensing performance of the FIM. For comparison, we consider four different benchmark transmission schemes:
\begin{itemize}
	\item \textbf{RAA-PA:} A conventional rigid phased array is utilized as the transmitter \cite{fuchsdr}.
	\item \textbf{FIM-PA:} An FIM-based phased array is utilized as the transmitter.
	\item \textbf{RAA-MIMO:} A conventional rigid MIMO array is utilized as the transmitter \cite{stoica2007prob}.
	\item \textbf{FIM-MIMO:} An FIM-based MIMO array is utilized as the transmitter.
\end{itemize}
Note that the rigid antenna array also corresponds to the scenario of integrating rigid RIS with the transceiver. The maximum morphing range of the FIM is set to $d_{\max }^t=\lambda$,\footnote{Note that the maximum morphing range of FIM relies on the array size. According to the latest research on the FIM prototype, 30\% out-of-plane deformation with respect to the array size can be achieved \cite{bai2022dynamically}. In the considered $10\times10$ FIM setup, this corresponds to $0.3\times9\times 0.5\lambda=1.35\lambda$.} and the transmit power is $P_t=10$ dBm. Since the MIMO array waveform design problem is the semi-definite relaxation (SDR) of the corresponding phased array beampattern design problem \cite{stoica2007prob}, we can readily modify problem \eqref{prob1} for the case of phased arrays by adding the constraint $\text{rank}\left({\bf{R}}_\text{X}\right)=1$, and the low-rank solution can be obtained by employing Gaussian randomization \cite{sidro2006mul}. Fig. \ref{bp} presents the beampatterns under the four considered transmission strategies, highlighting the significant enhancement in signal power at each target location and total signal power achieved through the dynamic surface shape morphing of the FIM. For the phased array, the minimum signal power at targets increases from $23.77$ dBm to $24.61$ dBm, while for the MIMO array, it increases from $23.56$ dBm to $26.64$ dBm. Due to the fact that only a single type of waveform is utilized in phased array systems, it has limited design DoF. Therefore, the total signal power it can achieve is lower compared to the MIMO array. This again demonstrates the advantage of MIMO arrays in maximizing power distribution and enhancing wireless sensing performance when combined with the flexibility of the FIM.
\begin{figure*}[]
	\centering
	\subfigure[]{
		\label{conv}
		\includegraphics[width=0.31\textwidth]{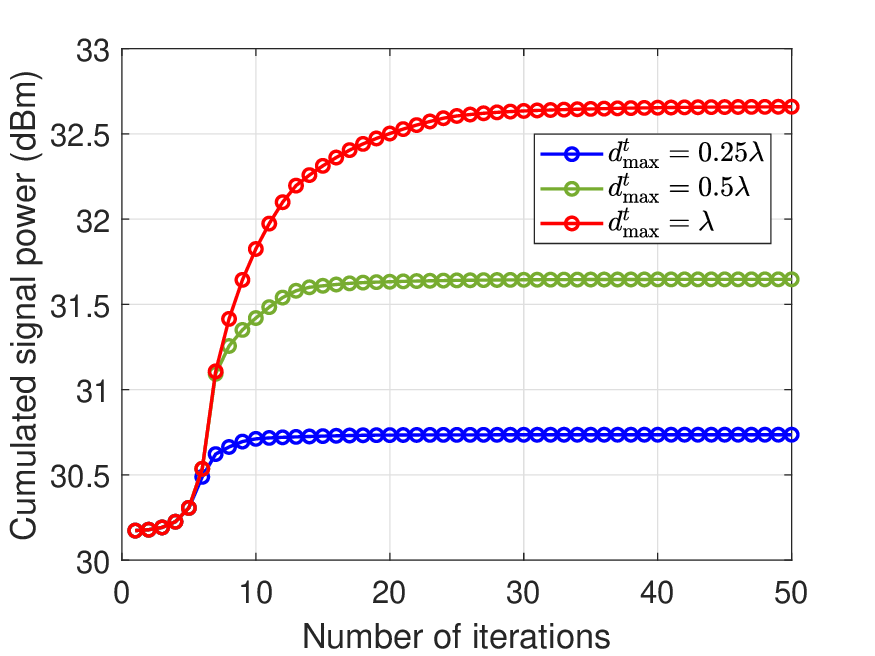}
	}
	\subfigure[]{
		\label{powdbm}
		\includegraphics[width=0.31\textwidth]{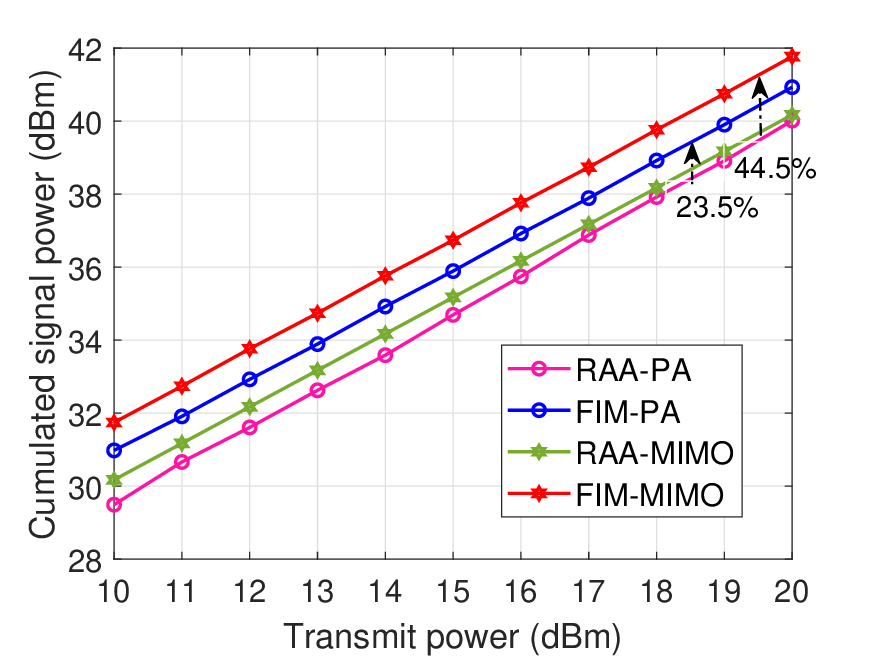}
	}
	\subfigure[]{
		\label{range}
		\includegraphics[width=0.31\textwidth]{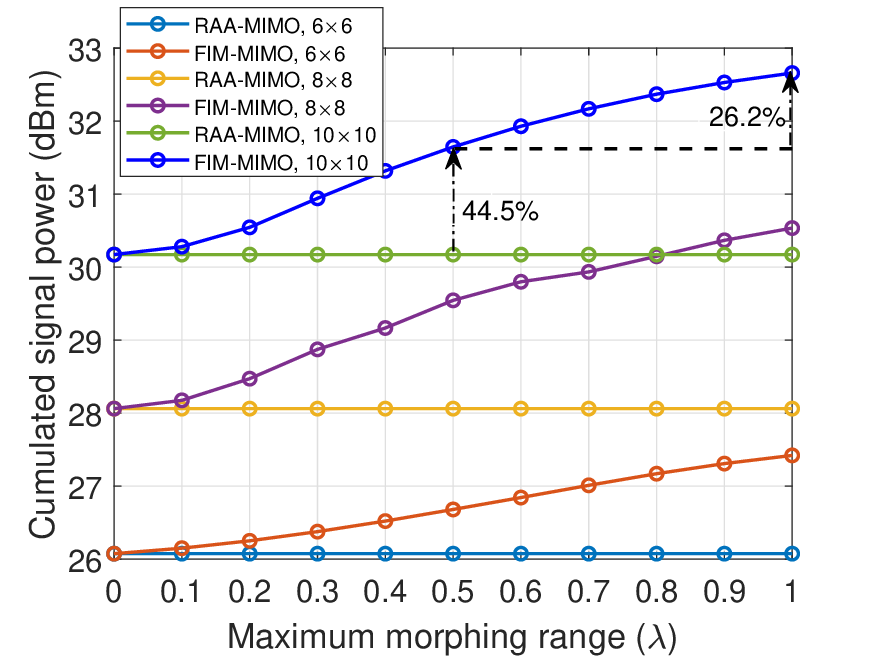}
	}
	\caption{(a) The convergence behavior of the proposed BCD algorithm; (b) Cumulated power at the target locations versus transmit power; (c) Cumulated power at the target locations versus maximum morphing range.}
\end{figure*} 

Furthermore, Fig. \ref{conv} illustrates the convergence curves of the proposed BCD algorithm. It can be observed that under different maximum morphing ranges, the algorithm achieves convergence within less than 35 iterations. Specifically, the smaller the maximum morphing range, the faster the proposed algorithm converges.

\subsection{Cumulated Power at the Target Locations Versus Transmit Power}
In this subsection, we evaluate the cumulated power at the target locations of these schemes versus the transmit power, where the maximum morphing range of the FIM is set to $d_{\max }^t=0.5\lambda$. As it can be seen from Fig. \ref{powdbm}, the cumulated power at the target locations of all schemes increases with the transmit power. Among all benchmark schemes, FIM-MIMO achieves the highest cumulated power at the target locations by flexibly morphing its surface shape to achieve favorable array steering vectors. Specifically, compared to the rigid array, FIM-PA and FIM-MIMO increase the cumulated power at the target locations by $23.5$\% and $44.5$\%, respectively.

\subsection{Cumulated Power at the Target Locations Versus Morphing Range}
In this subsection, we analyze the relationship between the cumulated power at the target locations and the FIM morphing range. The transmit power is set to $P_t=10$ dBm. As shown in Fig. \ref{range}, under the same maximum morphing range, the cumulated power at the target locations increases with the number of elements, and the cumulated power at the target locations increases significantly as the morphing range increases from $0$ to $0.5\lambda$, with an increase rate of $44.5$\% for a $10\times10$ array. However, when the morphing range is further increased from $0.5\lambda$ to $\lambda$, the cumulated power at the target locations exhibits a slower growth rate, with an increase rate of $26.2$\% for a $10\times10$ array. This indicates that while increasing the morphing range initially provides substantial gains, diminishing returns are observed as the morphing range continues to increase beyond $0.5\lambda$.

\section{Conclusion}
In this paper, we proposed an FIM-aided MIMO wireless sensing system, where the FIM dynamically morphs its surface shape to enhance sensing performance by increasing the cumulated signal power at target locations. To this end, we formulated a joint optimization problem involving the transmit covariance matrix and the surface shape of the transmitting FIM and developed an efficient BCD algorithm that iteratively optimize both variables. Simulation results demonstrated the enhancement of wireless sensing performance achieved by FIM. Specifically, when the maximum morphing range reaches half a wavelength, the FIM achieves more than 40\% performance improvement in terms of cumulated probing signal power over the conventional rigid MIMO array. In future work, we aim to extend our framework to account for dynamic target motion and time-varying channels, potentially by incorporating statistical models or robust optimization techniques to mitigate the impact of temporal decorrelation.

\begin{appendices}
	\section{The Gradient of the Objective Function in \eqref{rxgivenR} \\ \emph{w.r.t.} $\Delta {{\bf{d}}^t}$}	
	Firstly, according to \cite{2007-diff}, the partial derivative of $P_c$ \emph{w.r.t.} the deformation distance $\Delta d_{{n_t}}^t$ of the $n_t$-th transmitting element is given by
	\begin{align}\label{B1}
		\frac{\partial }{{\partial \Delta d_{{n_t}}^t}}\text{tr}\left({\bf{R}}_\text{X}{\bf{B}}\right)
		= \text{tr}\left({\bf{R}}_\text{X}\frac{\partial }{{\partial \Delta d_{{n_t}}^t}}{\bf{B}} \right).
	\end{align}
	Recall \eqref{BB} and by using the product rule in calculus $d\left( {{{\bf{Z}}_1}{{\bf{Z}}_2}} \right) = d\left( {{{\bf{Z}}_1}} \right){{\bf{Z}}_2} + {{\bf{Z}}_1}d\left( {{{\bf{Z}}_2}} \right)$, we have
	\begin{align}\label{BBB}
		\frac{\partial }{{\partial \Delta d_{{n_t}}^t}}{\bf{B}}
		= \left(\frac{\partial }{{\partial \Delta d_{{n_t}}^t}}{\bf{A}}\right){\bf{A}}^H+{\bf{A}}\left(\frac{\partial }{{\partial \Delta d_{{n_t}}^t}}{\bf{A}}\right)^H,
	\end{align}
	and the partial derivative $\frac{\partial }{{\partial \Delta d_n^t}}{\bf{A}}$ can be calculated as
	\begin{align}\label{PA}
		\frac{\partial }{{\partial \Delta d_{n_t}^t}}{\bf{A}} = - j{k_c}{\bf{\tilde A}}_{n_t}{\text{diag}}({{\bf{s}}_{\bf{\Theta }}} \odot {{\bf{s}}_{\bf{\Phi }}}),
	\end{align}
	where
		${{\bf{\tilde A}}_{{n_t}}} = \left[ {{{\bf{0}}_{({n_t} - 1) \times K}};{{\bf{A}}_{{n_t},:}};{{\bf{0}}_{({N_t} - {n_t}) \times K}}} \right] \in \mathbb{C}{^{{N_t} \times K}}$,
		${{\bf{s}}_{\bf{\Theta }}} = {\left[\sin {\theta _1},\sin {\theta _2}...,\sin {\theta _K}\right]^T}\in \mathbb{R}^{K \times 1}$,
		${{\bf{s}}_{\bf{\Phi }}} = {\left[\sin {\varphi _1},\sin {\varphi _2}...,\sin {\varphi _K}\right]^T}\in \mathbb{R}^{K \times 1}$,
	and ${{\bf{A}}_{{n_t},:}}$ denotes the $n_t$-th row of ${\bf{A}}$. By substituting \eqref{PA} and \eqref{BBB} into \eqref{B1}, we can get the partial derivative of $P_c$ \emph{w.r.t.} the deformation distance $\Delta d_{{n_t}}^t$.
	Finally, by collecting all $N_t$ partial derivatives, the gradient of $P_c$ \emph{w.r.t.} $\Delta {{\bf{d}}^t}$ is given by
	$	\nabla_{\Delta {{\bf{d}}^t}}P_c\
		=\left[\frac{\partial }{{\partial \Delta d_{1}^t}}\text{tr}\left( {\bf{R}}_\text{X}{\bf{B}} \right),\frac{\partial }{{\partial \Delta d_{2}^t}}\text{tr}\left( {\bf{R}}_\text{X}{\bf{B}} \right),\cdots,\frac{\partial }{{\partial \Delta d_{{N_t}}^t}}\text{tr}\left({\bf{R}}_\text{X}{\bf{B}} \right)\right]^T.$
\end{appendices}
\bibliographystyle{IEEEtran}
\bibliography{Reference}

\begin{thebibliography}{10}
\providecommand{\url}[1]{#1}
\csname url@samestyle\endcsname
\providecommand{\newblock}{\relax}
\providecommand{\bibinfo}[2]{#2}
\providecommand{\BIBentrySTDinterwordspacing}{\spaceskip=0pt\relax}
\providecommand{\BIBentryALTinterwordstretchfactor}{4}
\providecommand{\BIBentryALTinterwordspacing}{\spaceskip=\fontdimen2\font plus
\BIBentryALTinterwordstretchfactor\fontdimen3\font minus \fontdimen4\font\relax}
\providecommand{\BIBforeignlanguage}[2]{{%
\expandafter\ifx\csname l@#1\endcsname\relax
\typeout{** WARNING: IEEEtran.bst: No hyphenation pattern has been}%
\typeout{** loaded for the language `#1'. Using the pattern for}%
\typeout{** the default language instead.}%
\else
\language=\csname l@#1\endcsname
\fi
#2}}
\providecommand{\BIBdecl}{\relax}
\BIBdecl

\bibitem{you2021towards}
X.~You, C.-X. Wang, J.~Huang, X.~Gao, Z.~Zhang, S.~Hu, Y.~Jiang, J.~Wang, and L.~Yang, ``Towards {6G} wireless communication networks: Vision, enabling technologies, and new paradigm shifts,'' \emph{Sci. China Inf. Sci.}, vol.~64, pp. 1--74, Nov. 2021.

\bibitem{begum2024auto}
M.~Begum, G.~Raja, and M.~Guizani, ``{AI}-based sensor attack detection and classification for autonomous vehicles in {6G-V2X} environment,'' \emph{IEEE Trans. Veh. Technol.}, vol.~73, no.~4, pp. 5054--5063, Apr. 2024.

\bibitem{SP_2025_Teng_Frequency}
Z.~Teng \emph{et~al.}, ``Frequency invariant beamformer design exploiting {SRV}-constrained array response control,'' \emph{Signal Process.}, p. 110154, 2025.

\bibitem{IoTJ_2025_Teng_Dynamic}
------, ``Dynamic precoding for near-field secure communications: Implementation and performance analysis,'' \emph{IEEE Int. Things J.}, pp. 1--14, 2025, Early Access.

\bibitem{di2020smart}
M.~Di~Renzo, A.~Zappone, M.~Debbah, M.-S. Alouini, C.~Yuen, J.~de~Rosny, and S.~Tretyakov, ``Smart radio environments empowered by reconfigurable intelligent surfaces: How it works, state of research, and the road ahead,'' \emph{IEEE J. Sel. Areas Commun.}, vol.~38, no.~11, pp. 2450--2525, Nov. 2020.

\bibitem{an2024codebook}
J.~An, C.~Xu, Q.~Wu, D.~W.~K. Ng, M.~Di~Renzo, C.~Yuen, and L.~Hanzo, ``Codebook-based solutions for reconfigurable intelligent surfaces and their open challenges,'' \emph{IEEE Wireless Commun.}, vol.~31, no.~2, pp. 134--141, Apr. 2024.

\bibitem{buz2022fon}
S.~Buzzi, E.~Grossi, M.~Lops, and L.~Venturino, ``Foundations of {MIMO} radar detection aided by reconfigurable intelligent surfaces,'' \emph{IEEE Trans. Signal Process.}, vol.~70, pp. 1749--1763, Mar. 2022.

\bibitem{song2023crb}
X.~Song, J.~Xu, F.~Liu, T.~X. Han, and Y.~C. Eldar, ``Intelligent reflecting surface enabled sensing: {Cramér-Rao} bound optimization,'' \emph{IEEE Trans. Signal Process.}, vol.~71, pp. 2011--2026, May 2023.

\bibitem{lv2024noma}
W.~Lyu, Y.~Xiu, X.~Li, S.~Yang, P.~L. Yeoh, Y.~Li, and Z.~Zhang, ``Hybrid {NOMA} assisted integrated sensing and communication via {RIS},'' \emph{IEEE Trans. Veh. Technol.}, vol.~73, no.~5, May 2024.

\bibitem{wangwcnc2024}
D.~Wang, A.~Bazzi, and M.~Chafii, ``{RIS}-enabled integrated sensing and communication for {6G} systems,'' in \emph{Proc. IEEE Wireless Commun. Netw. Conf. (WCNC)}, 2024, pp. 1--6.

\bibitem{baz2025jsac}
A.~Bazzi and M.~Chafii, ``Low dynamic range for {RIS}-aided bistatic integrated sensing and communication,'' \emph{IEEE J. Sel. Areas Commun.}, vol.~43, no.~3, pp. 912--927, 2025.

\bibitem{TAP_2025_An_Emerging}
J.~An, M.~Debbah, T.~J. Cui, Z.~N. Chen, and C.~Yuen, ``Emerging technologies in intelligent metasurfaces: Shaping the future of wireless communications,'' \emph{IEEE Trans. Antennas Propag.}, pp. 1--16, 2025, Early Access.

\bibitem{an2023stacked}
J.~An, C.~Yuen, C.~Xu, H.~Li, D.~W.~K. Ng, M.~Di~Renzo, M.~Debbah, and L.~Hanzo, ``Stacked intelligent metasurface-aided {MIMO} transceiver design,'' \emph{IEEE Wireless Commun.}, vol.~31, no.~4, pp. 123--131, Aug. 2024.

\bibitem{niu2024isac}
H.~Niu, J.~An, A.~Papazafeiropoulos, L.~Gan, S.~Chatzinotas, and M.~Debbah, ``Stacked intelligent metasurfaces for integrated sensing and communications,'' \emph{IEEE Wireless Commun. Lett.}, vol.~13, no.~10, pp. 2807--2811, Oct. 2024.

\bibitem{an2023stacked2}
J.~An, C.~Xu, D.~W.~K. Ng, G.~C. Alexandropoulos, C.~Huang, C.~Yuen, and L.~Hanzo, ``Stacked intelligent metasurfaces for efficient holographic {MIMO} communications in {6G},'' \emph{IEEE J. Sel. Areas Commun.}, vol.~41, no.~8, pp. 2380--2396, Aug. 2023.

\bibitem{an2024two}
J.~An, C.~Yuen, Y.~L. Guan, M.~Di~Renzo, M.~Debbah, and L.~Hanzo, ``Two-dimensional direction-of-arrival estimation using stacked intelligent metasurfaces,'' \emph{IEEE J. Sel. Areas Commun.}, vol.~42, no.~10, pp. 2786--2802, Oct. 2024.

\bibitem{kamali2016decoupling}
S.~M. Kamali, A.~Arbabi, E.~Arbabi, Y.~Horie, and A.~Faraon, ``Decoupling optical function and geometrical form using conformal flexible dielectric metasurfaces,'' \emph{Nat. Commun.}, vol.~7, no.~1, p. 11618, May 2016.

\bibitem{bai2022dynamically}
Y.~Bai, H.~Wang, Y.~Xue, Y.~Pan, J.-T. Kim, X.~Ni, T.-L. Liu, Y.~Yang, M.~Han, and Y.~Huang, ``A dynamically reprogrammable surface with self-evolving shape morphing,'' \emph{Nature}, vol. 609, no. 7928, pp. 701--708, Sep. 2022.

\bibitem{anfim2025}
J.~An, C.~Yuen, M.~D. Renzo, M.~Debbah, H.~Vincent~Poor, and L.~Hanzo, ``Flexible intelligent metasurfaces for downlink multiuser {MISO} communications,'' \emph{IEEE Trans. Wireless Commun.}, 2025, {E}arly {A}ccess.

\bibitem{arXiv_2025_An_Flexible}
J.~An, Z.~Han, D.~Niyato, M.~Debbah, C.~Yuen, and L.~Hanzo, ``Flexible intelligent metasurfaces for enhancing {MIMO} communications,'' \emph{IEEE Trans. Commun.}, pp. 1--1, 2025, Early Access.

\bibitem{zhou2024fluid}
L.~Zhou, J.~Yao, M.~Jin, T.~Wu, and K.-K. Wong, ``Fluid antenna-assisted {ISAC} systems,'' \emph{IEEE Wireless Commun. Lett.}, vol.~13, no.~12, pp. 3533--3537, Dec. 2024.

\bibitem{ma2024ma}
W.~Ma, L.~Zhu, and R.~Zhang, ``Movable antenna enhanced wireless sensing via antenna position optimization,'' \emph{IEEE Trans. Wireless Commun.}, vol.~23, no.~11, pp. 16\,575--16\,589, Nov. 2024.

\bibitem{zhang2015duplex}
Z.~Zhang, X.~Chai, K.~Long, A.~V. Vasilakos, and L.~Hanzo, ``Full duplex techniques for {5G} networks: self-interference cancellation, protocol design, and relay selection,'' \emph{IEEE Commun. Mag.}, vol.~53, no.~5, pp. 128--137, May 2015.

\bibitem{stoica2007prob}
P.~Stoica, J.~Li, and Y.~Xie, ``On probing signal design for {MIMO} radar,'' \emph{IEEE Trans. Signal Process.}, vol.~55, no.~8, pp. 4151--4161, Aug. 2007.

\bibitem{tsp-2008-li-crb}
J.~Li, L.~Xu, P.~Stoica, K.~W. Forsythe, and D.~W. Bliss, ``Range compression and waveform optimization for {MIMO} radar: A {Cramér–Rao} bound based study,'' \emph{IEEE Trans. Signal Process.}, vol.~56, no.~1, pp. 218--232, Jan. 2008.

\bibitem{cvx}
M.~Grant and S.~Boyd, ``{CVX}: Matlab software for disciplined convex programming, version 2.1,'' Mar. 2014, [Online]. Available: \url{http://cvxr.com/cvx}.

\bibitem{2010_sdp_luo}
Z.-Q. Luo, W.-K. Ma, A.~M.-C. So, Y.~Ye, and S.~Zhang, ``Semidefinite relaxation of quadratic optimization problems,'' \emph{IEEE Signal Process. Mag.}, vol.~27, no.~3, pp. 20--34, May 2010.

\bibitem{fuchsdr}
B.~Fuchs, ``Application of convex relaxation to array synthesis problems,'' \emph{IEEE Trans. Antennas Propag.}, vol.~62, no.~2, pp. 634--640, Feb. 2014.

\bibitem{sidro2006mul}
N.~Sidiropoulos, T.~Davidson, and Z.-Q. Luo, ``Transmit beamforming for physical-layer multicasting,'' \emph{IEEE Trans. Signal Process.}, vol.~54, no.~6, pp. 2239--2251, Jun. 2006.

\bibitem{2007-diff}
A.~Hjorungnes and D.~Gesbert, ``Complex-valued matrix differentiation: Techniques and key results,'' \emph{IEEE Trans. Signal Process.}, vol.~55, no.~6, pp. 2740--2746, Jun. 2007.

\end{thebibliography}
\end{document}